%% file: main.tex
\documentclass[sigconf]{acmart}

\AtBeginDocument{%
  \providecommand\BibTeX{{%
    \normalfont B\kern-0.5em{\scshape i\kern-0.25em b}\kern-0.8em\TeX}}}

\setcopyright{none} 
\copyrightyear{}
\acmYear{2024}
\acmDOI{} 

\acmConference[CHI'24 WS 25]{ACM CHI 2024 Workshop WS 25: Designing Inclusive Future Augmented Realities}{May 12, 2024}{Honolulu, Hawai'i}

\acmBooktitle{ACM CHI 2024 Workshop WS 25: Designing Inclusive Future Augmented Realities, May 12, 2024 in Honolulu, Hawai'i}
\acmPrice{}
\acmISBN{}

\usepackage{xspace}
\newcommand{\etal}[1]{{#1}~et~al.}
\newcommand{\ie}{i.e.,\xspace}

\newcommand{\eg}{e.g.,\xspace}

\begin{document}

    \title{Don't Leave Me Out: Designing for \\Device Inclusivity in Mixed Reality Collaboration}

    \input{content/_authors}
    \renewcommand{\shortauthors}{Krug et al.}


    \maketitle
    
    \input{content/1_motivation}
    \input{content/2_position}

    \input{content/3_conclusion}
    \input{content/4_author_info}

    \input{content/_outro}

    \bibliographystyle{ACM-Reference-Format}
    \bibliography{refs}
    
\end{document}

%% file: content/_authors.tex
\author{Katja Krug}
\authornote{Emails: \{katjakrug, dachselt\}@acm.org, \{julian.mendez2, weizhou.luo\}@tu-dresden.de}
\authornote{Also with the Centre for Tactile Internet with Human-in-the-Loop (CeTI), TU Dresden}
\orcid{0000-0003-4800-6287}
\affiliation{%
    \institution{Interactive Media Lab Dresden, TU Dresden}
  \country{Germany}
}

\author{Juli\'{a}n M\'{e}ndez}
\authornotemark[1]
\orcid{0000-0003-1029-7656}
\affiliation{
    \institution{Interactive Media Lab Dresden, TU Dresden}
    \country{Germany}
}

\author{Weizhou Luo}
\authornotemark[1]
\orcid{0000-0002-1312-1528}
\affiliation{
    \institution{Interactive Media Lab Dresden, TU Dresden}
    \country{Germany}
}

\author{Raimund Dachselt}
\authornotemark[1]
\authornotemark[2]
\orcid{0000-0002-2176-876X}
\affiliation{
    \institution{Interactive Media Lab Dresden, TU Dresden}
    \country{Germany}
}

%% file: content/1_motivation.tex

\section{Mixed Reality collaboration and economic exclusion}

Modern collaborative Mixed Reality (MR) systems continue to break the boundaries of conventional co-located and remote collaboration and communication. 
They merge physical and virtual worlds 
and enable natural interaction, opening up a spectrum of novel opportunities for interpersonal connection.
For these connections to be perceived as engaging and positive, collaborators should feel comfortable and experience a sense of belonging \cite{belonging}. 
Not having the dedicated devices to smoothly participate in these spaces can hinder this and give users the impression of being left out.
To counteract this, we propose to prioritize designing for device inclusivity in MR collaboration, focusing on compensating disadvantages of common non-immersive device classes in cross-device systems.   

Many MR experiences are mainly bound to expensive flagship 
Head-Mounted Displays (HMDs), with built-in capabilities like head-, eye-, and hand-tracking and the capacity for 3D holographic projection. 
As of today, it is estimated that less than 200 million users worldwide have access to dedicated Augmented or Virtual Reality (AR and VR) technology~\cite{zippiaAmazingVirtual}
, which pales in comparison to the approximately 5.4 billion users of smartphones~\cite{datareportalDigital2023}.
The tailoring of collaborative MR experiences to high-end HMDs therefore creates an implicit economic divide, excluding people from social experiences and interpersonal connection based on access to expensive and exclusive hardware. 
Moreover, even among high-end MR devices, the heterogeneity and low interoperability of current technologies confine many collaborative MR experiences to a very exclusive circle of users. 
Realistically, a group of collaborators brainstorming in an immersive environment probably won't each have access to the same high-end device, but various different technologies. 
To facilitate this kind of collaboration, we advocate for changing the mindset towards designing MR systems to \textit{prioritize inclusivity and accessibility} regardless of device class.
Toward this goal, the following questions arise:
How can MR systems be designed to offer a comparable experience on fundamentally different technologies? 
How can we mitigate the disadvantages of less powerful or non-dedicated devices, 
without undermining high-end MR devices?


While striving for accessibility, we want to avoid simply building around the lowest common denominator and failing to leverage the capabilities of high-end devices.
Instead, we draw inspiration from established adaptive design concepts, such as 
\textit{responsive web design}, which enables modern websites to adapt their layout and content depending on the device in which they are accessed. 
Another source of inspiration is the gaming scene. 
Games often support \textit{cross-platform} online gameplay,
along with a range of graphics settings to accommodate for the greatly varying computational power of the used devices.
By adapting some of these techniques for MR collaboration, we assume that a satisfying experience can be offered to users of flagship technology, without disadvantaging others.

While there is no shortage of concepts for asymmetric cross-device collaboration in MR systems, the prioritization of inclusivity is less prevalent. 
This might be due to asymmetric environments often implying asymmetric roles, where direct comparison between collaborators is less justified and a feeling of being left out less likely. 
Non-immersive devices, such as handhelds or desktops often assume an observing \cite{Cross:2022-AR-VR-MP-Tablet,Cross:VR+Tablet,Cross:MR+Tablet} or instructing role \cite{Cross:2023-VR-Laptop-Tablet-Smartphones}, provide guidance for HMD users \cite{Cross:Desktop+AR,Cross:Desktop+MRguidance,Cross:Phone+VR}, or perform other tasks that are distinctly different from the HMD users \cite{Cross:MR+Tablet,Cross:VR+HandheldTablet}. 
Here, additional non-immersive devices are employed in a complementary way towards a main actor or a common goal.
However, when different devices assume equal roles \cite{Cross:2017-AR+Tablet,Cross:2023-MR-VR+Desktop,Cross:AR+Phone,Cross:HMD+Phone+Projection} and collaborators are easily comparable to one another, the disparity in capabilities of non-dedicated devices is seldomly compensated to benefit users directly.

In this work, we want to advocate for designing truly device inclusive MR collaboration systems, shifting the mindset from asking "what can non-immersive devices do for the higher-end actors?", to \textit{"what can we do to accommodate and support lower-end devices as equally important actors?"}
We want to look at these platforms from a different perspective, by designing spaces that allow users to feel equal regardless of financial means or technical equipment, and create a sense of belonging. 

%% file: content/2_position.tex
\section{Considerations for device inclusive Mixed Reality applications}
An optimally designed collaborative cross-device MR system should make users feel like they equally belong to the collaborative space regardless of the device they use. 
In this context, equality does not necessarily mean equity. 
Providing the same tools to non-immersive users might not result in them feeling equally empowered when sharing a space with high-end HMD users. 
Many MR applications are currently developed prioritizing the top of the device hierarchy, primarily focusing on leveraging all capabilities of dedicated high-end HMDs, and then introducing additional devices into the already established space. 
We propose an iterative design approach, where the highest and lowest end of the hierarchy are equally considered for each feature of the application.
While developing features for the top, we suggest to consider in parallel the design of technological and conceptual components around the capabilities at the bottom, focusing on compensating the disadvantages in comparison to the devices at the top.
Moving upwards along the hierarchy, these compensations can be gradually adapted or removed, in accordance with the increasing capabilities of higher-end devices.
With this, we aim to shatter the perception of high-end HMDs as the "default device" and encourage to think about collaborative MR spaces as an ecosystem of different devices.
By enabling intelligent content adaptation based on device capabilities, we can attempt to model a form of "responsiveness" in MR applications.

\subsection{Technological considerations}
When attempting to prioritize inclusivity in collaborative MR systems, we need to consider how applications have to be designed to offer consistent experiences on various devices independent of computing or rendering power. 
Many expensive tasks can be offloaded to external servers and streamed to the devices, if the network is sufficient.
Power demand can be reduced on lower-tech devices by employing techniques that reduce the graphics workload, such as loading objects only when within view (\ie lazy loading \cite{Techniques:LazyLoading}), using less demanding or less detailed representations of objects (\ie adaptive level-of-detail \cite{Techniques:AdaptiveLOD}), or reconstruction techniques to increase performance (\eg Unreal Engine's temporal upscalers~\cite{unrealengineTemporalUpscalers}).

\subsection{Design considerations}
We imagine disadvantage compensations for non-immersive devices in the form of awareness accommodations and additional interactive functionality, to level out discrepancies between device classes. 
Here, we can draw inspiration from team-based and asymmetric multiplayer video-games where characters have different abilities and roles which rely on \textit{balancing} adjustments so that their participation has a comparable influence on the game.

In the following, we want to illustrate some specific examples against the backdrop of a collaborative AR space, consisting of high-end HMD users and handheld mobile AR users, who perceive the scene through smartphone screens.  

One way to support disadvantaged devices could be to bring awareness about potential shortcomings to HMD collaborators.
For example, a 3D visualization of a viewing cone attached to the smartphone could potentially help HMD users to consider the current Field of View (FoV) of the smartphone user, and encourage HMD users to make them aware of objects of interest outside of their FoV, similar to what is implemented by \etal{Müller} \cite{Cross:2019-TabletAR+VR} in their symmetric tablet collaboration.
Similarily, the visualization of 3D pointing rays or cursors following the path of touch input on a smartphone could enable HMD users to easily identify referenced objects in the scene, a concept which is similarily implemented by \etal{Norman} \cite{Cross:Desktop+AR}.

Additionally, the smartphone user could be directly empowered through awareness cues that typical non-immersive setups don't provide, such as peripheral visual perception and spatial audio.
Here, traditional off-screen visualizations come to mind, such as indications on the borders of the screen about the position of objects of interest, approaching virtual collaborators or which direction a sound is coming from. 

Besides awareness accommodations, there are also major discrepancies regarding interaction possibilities.
Naturally, smartphone users need to have sufficient alternatives to direct mid-air gesture interaction.
For object manipulation or selection, touch input can often be a sufficient substitution, as shown by \etal{Grandi} \cite{Cross:AR+Phone} and \etal{Speicher} \cite{Cross:HMD+Phone+Projection} among others.
Smartphone users could also possess additional powers, like the ability to decouple themselves from the virtual space by allowing them to zoom out of the scene and gain a bird's-eye view. 
These powers need to be carefully considered, as they could be perceived as an unfair benefit by HMD users, leading to discontentment on the basis of perceived favoritism.
The challenge is to allow for enough additional power to empower the disadvantaged, without discriminating against the privileged.

%% file: content/3_conclusion.tex
\subsection{Conclusion}
In summary, we believe that prioritizing device inclusivity while designing  collaborative cross-device MR platforms bears great potential to make MR spaces more accessible for a bigger audience. 
We propose the integration of disadvantage compensations and point out, that they need to be carefully considered to empower, but not overpower individual users, and convey an equal sense of belonging to the collaborative space. 
We call for the creation of unified platforms that allow dynamic collaborations among different devices and device classes, and we are looking forward to fruitful discussions about inclusivity in future Mixed Realities. 

%% file: content/4_author_info.tex

\section{Authors and Affiliations}

All authors are members (and head) of the Interactive Media Lab Dresden (IMLD). 
\textbf{\href{https://imld.de/en/our-group/team/katja-krug/}{Katja Krug}} is a PhD student that develops and researches collaborative Mixed Reality applications and explores interaction with devices for 3D in- and output.
\textbf{\href{https://imld.de/en/our-group/team/julian-mendez/}{Juli\'an M\'endez}} is a PhD student that develops interactive visualization approaches for exploration, analysis and explanation of computational models since 2017.
\textbf{\href{https://imld.de/en/our-group/team/weizhou-luo/}{Weizhou Luo}} is a fifth year PhD student, and his research revolves around applying Mixed and Augmented Reality to support data exploration, interaction, and sensemaking.
\textbf{\href{https://imld.de/en/our-group/team/raimund-dachselt/}{Raimund Dachselt}}
is the head of the IMLD since 2012 and director of the Institute of Software and Multimedia Technology at TU Dresden since 2015.
His and his team's research is focused on interactive data visualization, interactive surfaces, multimodal human-computer interaction, physical computing, and Mixed Reality interfaces.

%% file: content/_outro.tex
\begin{acks}
This work was supported by the Deutsche Forschungsgemeinschaft (DFG, German Research Foundation) under Germany’s Excellence Strategy: 
EXC-2068, 390729961 – Cluster of Excellence “Physics of Life” and
EXC 2050/1, 390696704 – Cluster of Excellence “Centre for Tactile Internet” (CeTI) of TU Dresden, by DFG grant 389792660 as part of TRR~248 -- CPEC (see \url{https://cpec.science}) and by the German Federal Ministry of Education and Research (BMBF, SCADS22B) and the Saxon State Ministry for Science, Culture and Tourism (SMWK) by funding the competence center for Big Data and AI "ScaDS.AI Dresden/Leipzig". 
\end{acks}